\documentclass[reprint, nofootinbib, pra]{revtex4} 
\usepackage{amssymb, amsmath}
\usepackage{graphicx}
\usepackage{verbatim}
\usepackage{xcolor}
\begin{document}
\title{Analysis of the superdeterministic Invariant-set theory in a hidden-variable setting}
\author{Indrajit Sen}
\email{isen@chapman.edu}
\affiliation{Institute for Quantum Studies, Chapman University\\
One University Drive, Orange, CA, 92866, USA}
\date{\today}
\begin{abstract}
A recent proposal for a superdeterministic account of quantum mechanics, named Invariant-set theory, appears to bring ideas from several diverse fields like chaos theory, number theory and dynamical systems to quantum foundations. However, a clear cut hidden-variable model has not been developed, which makes it difficult to assess the proposal from a quantum foundational perspective. In this article, we first build a hidden-variable model based on the proposal, and then critically analyse several aspects of the proposal using the model. We show that several arguments related to counter-factual measurements, nonlocality, non-commutativity of quantum observables, measurement independence etcetera that appear to work in the proposal fail when considered in our model. We further show that our model is not only superdeterministic but also nonlocal, with an ontic quantum state. We argue that the bit string defined in the model is a hidden variable and that it contains redundant information. Lastly, we apply the analysis developed in a previous work (Proc. R. Soc. A, 476(2243):20200214, 2020) to illustrate the issue of superdeterministic conspiracy in the model. Our results lend further support to the view that superdeterminism is unlikely to solve the puzzle posed by the Bell correlations.
\end{abstract}
\maketitle

\section{Introduction}
Bell's theorem \cite{bell} continues to challenge our understanding of the relationship between the two pillars of modern physics: quantum mechanics and relativity. In order to circumvent the theorem, a realist interpretation of quantum mechanics must violate at least one of its assumptions\footnote{For example, pilot-wave theory \cite{louis27, solventini, bohm1, bohm2} violates the local-causality assumption whereas the many-worlds interpretation \cite{everest, sepmolds} violates the single-outcome assumption.}. In recent years, the measurement-independence assumption\footnote{Sometimes also referred to as statistical-independence \cite{wein09, sabinethink} or $\lambda$-independence \cite{whartreview}.} in Bell's theorem has received significant attention in the literature \cite{wein09, hall10, howmuch, hall16, cosmicbellII, hall19}. The assumption states that the hidden variables that determine the measurement outcomes are uncorrelated with the measurement settings. Several models have been developed that violate this assumption and thereby circumvent the theorem (for a recent survey of such models, see ref. \cite{whartreview}). It has also been possible to exploit the properties of these models to incorporate relativistic effects on entangled quantum systems \cite{seneffect}.  However, no wide consensus has yet emerged on how to physically interpret the violation of measurement independence.\\

There are, at present, two options for a physical interpretation of the violation of measurement independence: retrocausality and superdeterminism. Retrocausality \cite{costacoffee, cramer, pricebook, sutherland, whartonmain, lazar15, fpaper, whartreview} is the idea that events are not fully determined by past conditions alone, but that future conditions must be specified as well. Retrocausal models are, therefore, not deterministic given only the past conditions. The information about the measurement settings are encoded in the future boundary conditions. This information is then thought to, in some sense, causally influence the hidden-variable distribution at the time of preparation backwards in time. This results in violation of the measurement-independence assumption. How to reconcile our intuitive understanding of causation and time with retrocausality is a major conceptual question for this approach. The other option is superdeterminism \cite{dialect}. Unlike retrocausal models, superdeterministic models (for examples, see ref's. \cite{brans, hooft16}) are deterministic given the past conditions. Specifically, not only are the measurement outcomes determined by the past conditions, but the measurement settings as well. The measurement-independence assumption is violated by positing that the initial conditions enforce a correlation between the hidden variables and the measurement settings. How to justify such initial conditions is a major conceptual question for this approach. It has been argued in ref's. \cite{1st, 2nd} that any such justification would necessarily be conspiratorial in a quantitative sense.\\

Palmer has been proposing, in increasing detail over several years, a superdeterministic account of quantum mechanics named Invariant-set theory \cite{palmer09, palmer1, palmer2, palmerend, palmerpost}. His proposal (for convenience, referred to as ``the Proposal'' hereafter) combines ideas from several diverse fields. The Proposal also gives a novel justification for the choice of initial conditions in terms of state-space geometry. However, unless the Proposal is condensed into a hidden-variable model, many details are bound to remain unclear from a quantum foundational perspective. A concrete model is also required in order to clarify to what extent the properties claimed by the Proposal actually hold up. In this article, we attempt to fill this gap by constructing a hidden-variable model based on the Proposal. We show that the resulting model is not only superdeterministic, but also nonlocal and $\psi$-ontic \cite{harrikens, pbr, leifer}. We also show how several arguments made in the Proposal fail when considered in the model. Lastly, we use our model and recent results about superdeterministic conspiracy \cite{1st, 2nd} to discuss the conspiratorial nature of superdeterminism in the Proposal.\\

To build our model, we consider the latest version of the Proposal, given in ref. \cite{palmerend}. The present article is structured as follows. We first give a brief, intuitive sketch of the Proposal in section \ref{r1}. We then build a hidden-variable model for single spin-1/2 particles based on the Proposal in section \ref{r2}. We use this model to analyse the various arguments about single-particle measurements made by the Proposal in section \ref{r3}. We extend the model to the Bell scenario in section \ref{r4}, and use the extended model to analyse several arguments about the Bell scenario made in the Proposal in section \ref{r5}. We also discuss the conspiratorial nature of superdeterminism in the model in \ref{r6}. We conclude with a discussion of our results in section \ref{r7}.

\section{Brief sketch} \label{r1}
In any realistic scenario, the experimentally-selected setting of a measurement apparatus (say the orientation of a Stern Gerlach) is different from its exact setting due to various errors. Due to the finite resolution of the apparatuses used for the setup, these errors cannot be completely eliminated. The Proposal considers the exact setting of an apparatus to be objectively real, unlike a `Copenhagenish' viewpoint that may deny the reality of any variable that cannot be operationally measured. It also considers the exact setting to be a fundamentally uncontrollable and unknowable quantity that is continuously varying with time. \\

In the Proposal, the exact setting of an apparatus may depend on the past exact setting of another apparatus at an arbitrary distance. This is due to certain rationality constraints that the Proposal imposes on the exact orientations of apparatuses involved in a quantum experiment. It considers these constraints to arise from the geometry of state space. Only those exact settings that satisfy these constraints are considered physically possible. \\

For these physically possible exact settings, a bit-string representation of the prepared quantum state can be constructed for the experiment. Each element in the bit string is a possible measurement outcome (for example, the elements will be $\pm1$'s for a single spin-1/2 particle). For a particular run of the experiment, one of the elements is selected, which determines the measurement outcome for that run. In a nutshell, these are the ideas that we will develop into a hidden-variable model. \\

The Proposal naturally suggests that we treat the uncontrollable, unknowable and continuously varying exact apparatus settings as hidden variables. It is difficult, however, to fix a particular value of the exact setting for each run in general. This is because the exact setting continuously varies with time and, for a physical measurement that takes a finite time, no single value of the exact setting can be specified. For example, the orientation of a Stern-Gerlach apparatus will continuously vary as a quantum particle passes through its magnetic field. For our purposes, we resolve this problem by considering ideal von-Neumann measurements \cite{vonN}. In these measurements, the apparatus is coupled to the quantum particle for a very short time $\delta t$ with a very high coupling constant $g$. The average exact setting over $\delta t$ will, in general, vary from one run to the next. Thereby, we can treat this average exact setting (over $\delta t$) as a hidden variable, with an associated hidden-variable distribution for an ensemble of runs.\\

Lastly, we suppose that there is a mechanism that takes the experimenter's choice of setting and orients the measuring apparatus accordingly. The final exact orientation of the apparatus depends on the experimenter's choices and the initial exact orientation of the apparatus. We are now ready to build a hidden-variable model based on the Proposal for spin-1/2 particles in the next section.\\

\section{The model for single spin-1/2 particles}\label{r2}
Consider a Stern-Gerlach measurement on an ensemble of unentangled spin-1/2 particles. Let the preparation procedure consist of measuring the ensemble of particles along a certain direction and then post-selecting only those particles that give the result +1/2. Let the experimenter select the initial orientation of the Stern Gerlach to be $\hat{p}$. We call $\hat{p}$ as the experimentally-selected orientation of the Stern Gerlach. Let the (uncontrollable and continuously varying) exact orientation of the Stern Gerlach be labelled by $\hat{P}(t)$. We label the experimental error $\hat{\delta}p(t) \equiv \hat{P}(t) - \hat{p}$. We assume that $|\hat{p}| = |\hat{P}(t)| = 1$ and $|\hat{\delta}p(t)| < \Delta$, where $\Delta$ is the minimum distance measurable by the finite-resolution apparatuses used to set up the experiment. This implies that the error $\hat{\delta}p(t)$ is, in principle, unmeasurable by the experimenters.\\

Let the experimenter choose the preparation setting to be $\hat{a}$ for a particular experimental run. We suppose that there is a mechanism that takes the experimenter's choice as input and orients the Stern Gerlach accordingly. Let the exact apparatus orientation when the mechanism begins orienting the apparatus to $\hat{a}$ be $\hat{P}$.  The final exact orientation of the apparatus (labelled by, say, $\hat{A}$) will, in general, depend on the initial exact orientation $\hat{P}$, the experimentally-chosen settings $\hat{p}$, $\hat{a}$ and other factors like frictional effects, various experimental parameters etc. We assume for simplicity that $\hat{A}$ can be expressed as the function $\hat{A} = \hat{A}\big (\hat{P}, \hat{p}, \hat{a} \big )$. It is straightforward to include other variables in the definition of the function $\hat{A}$, but this complicates the mathematical analysis without giving any physical insight. We also assume, for simplicity, that there is negligible delay between setting up the apparatus and performing the experiment, so that $\hat{A}\big (\hat{P}, \hat{p}, \hat{a} \big )$ is the actual setting used during the experiment for that run. Lastly, we assume that the function $\hat{A}$ is such that $|\hat{A} (\hat{P}, \hat{p}, \hat{a}) - \hat{a}| < \Delta$ $\forall \hat{P}, \hat{p}, \hat{a}$. This ensures that the difference between the exact and experimentally-selected apparatus orientations is unmeasurable by the experimenters.\\

We assume a similar process for setting up the Stern Gerlach at the measurement end. Let the experimenter select the initial orientation of the Stern Gerlach here to be $\hat{m}$. Suppose the experimenter select the measurement setting for a particular run to be $\hat{b}$. Let the exact apparatus orientation, when the mechanism begins orienting the apparatus to $\hat{b}$, be labelled by $\hat{M}$. We assume that $|\hat{m}| = |\hat{M}| =1$ and $|\hat{m} - \hat{M}| < \Delta$.  The initial exact orientation $\hat{M}$ is then mapped to the final exact orientation $\hat{B}$ such that the function $\hat{B} = \hat{B}(\hat{M}, \hat{m}, \hat{b})$ can be defined. We assume that the function $\hat{B}$ is such that $|\hat{\delta}b| \equiv |\hat{B}\big (\hat{M}, \hat{m}, \hat{b}\big ) - \hat{b}| < \Delta$ $\forall \hat{M}, \hat{m}, \hat{b}$.\\

According to the Proposal, only those pairs of $\hat{A}$ and $\hat{B}$ that satisfy the constraint
\begin{align}
&\hat{A} \cdot \hat{B} = 1 - \frac{2n-1}{N/2} \label{a}
\end{align}
for some $n \in \{1, 2,...N/2\}$ are considered to be physically possible, where $N$ is very large even constant. This does not, according to the Proposal, imply a causal relationship between $\hat{A}$ and $\hat{B}$. Instead, the initial conditions are assumed to be such that they lead to correlated pairs of  $\hat{A}$, $\hat{B}$ that satisfy (\ref{a}). Constraint (\ref{a}) implies that, given $\hat{A}$, $\hat{B}$ can occupy points only on a discrete set of points on the Bloch sphere. Suppose that, for a particular run, $\hat{A}\cdot \hat{B}$ satisfies the constraint (\ref{a}) for some $n$. We know that the quantum state of the particle can be expressed as 
\begin{align}
|+\rangle_{\hat{A}} = \cos (\theta_{AB}/2) |+\rangle_{\hat{B}} + e^{i\phi_{AB}} \sin (\theta_{AB}/2) |-\rangle_{\hat{B}} \label{gtf}
\end{align}
where $ \cos \theta_{AB} = \hat{A}\cdot \hat{B}$. The Proposal then defines a bit string that consists of $N\cos^2 (\theta_{AB}/2)$ +1 elements and $N\sin^2 (\theta_{AB}/2)$ -1 elements. Note that $N\cos^2 (\theta_{AB}/2)$ and $N\sin^2 (\theta_{AB}/2)$ are integers given the constraint (\ref{a}). Therefore, given $\hat{A}$, each point on the discretised Bloch sphere for $\hat{B}$ is associated with a particular bit string.\\ 

Let us label the bit string by a $1 \times N$ matrix $L_{1 \times N} (|+\rangle_{\hat{A}}, \hat{B})$. The exact ordering of the elements in $L_{1 \times N} (|+\rangle_{\hat{A}}, \hat{B})$ is determined by $\phi_{AB}$ in equation (\ref{gtf}). It is supposed that $\phi_{AB} = 0$ for the experiment, as the value of $\phi_{AB}$ depends on the purely theoretical choice of orientation of the axes, and does not affect the measurement results. However, according to the Proposal, if the particle is subsequently measured along a second direction $\hat{c}$ after the first experiment, then $\phi_{BC}$ will in general be non zero and will have to satisfy the constraint 
\begin{align}
\phi_{BC} = 2\pi l/N
\end{align} 
for some $l \in \{1, 2... N\}$. \\  

Given the bit string, the measurement outcome $O$ is determined in the following manner. We first define a variable $k \in \{1, 2,....N\}$ that determines the position on the bit string. The variable $k$ determines the particular `trajectory' (as described in the Proposal) out of $N$ possibilities that the quantum system can be in. Second, the Proposal specifies that the outcome is the $k^{th}$ element in the bit string $L_{1, k} (|+\rangle_{\hat{A}}, \hat{B})$.\\

We now switch to define the hidden variables in the model. Given the experimental choices $\hat{p}$, $\hat{a}$ ($\hat{m}$, $\hat{b}$) at the preparation (measurement) end, the measurement outcome is determined by a specification of the variables $\hat{P}$, $\hat{M}$ and $k$ (see Fig 1). That is, the measurement outcome can be expressed as $O\big (\hat{A}(\hat{P}, \hat{p}, \hat{a}), \hat{B}(\hat{M}, \hat{m}, \hat{b}), k\big ) = L_{1, k} \big (|+\rangle_{\hat{A}(\hat{P}, \hat{p}, \hat{a})}, \hat{B}(\hat{M}, \hat{m}, \hat{b}) \big )$. This implies that $\hat{P}$, $\hat{M}$, $k$ act as hidden variables in our model. We define $\mu \equiv (\hat{P}, \hat{M},  k)$. Note that the variables in $\mu$ are not causally affected by the experimental choices $\hat{a}$ and $\hat{b}$. This turns out to be important when considering a counter-factual change of $\hat{a}$, $\hat{b}$ for a particular run (as in section \ref{ola}, for example), as $\mu$ for the run does not change for the counter-factual case. \\

We can also directly express the measurement outcome as a function of the final exact apparatus settings $\hat{A}$, $\hat{B}$ and the position on the bit string $k$, or the exact quantum state prepared $|+\rangle_{\hat{A}}$, $\hat{B}$ and $k$, or the bit string $L_{1 \times N} (|+\rangle_{\hat{A}}, \hat{B})$ and $k$. The variables $\hat{A}$,  $|+\rangle_{\hat{A}}$, $\hat{B}$ and $L_{1 \times N} (|+\rangle_{\hat{A}}, \hat{B})$ are specified given the experimental choices $\hat{p}$, $\hat{a}$ ($\hat{m}$, $\hat{b}$) at the preparation (measurement) end and $\mu$. Note that the experimental choices are themselves determined by further hidden variables in a superdeterministic model. This implies that the variables $\hat{A}$,  $|+\rangle_{\hat{A}}$, $\hat{B}$ and $L_{1 \times N} (|+\rangle_{\hat{A}}, \hat{B})$ are determined by $\mu$ and the hidden variables that determine the experimental choices, and therefore have ontological status in our model. Lastly, the ontic state of the quantum system after the preparation is complete is given by $(|+\rangle_{\hat{A}}, k)$. \\

Let us describe the distribution of $\mu$ for an ensemble of runs. The Proposal interprets constraint (\ref{a}) to imply that the orientation $\hat{A}$ can be considered arbitrary but only those values of $\hat{B}$ that satisfy (\ref{a}) are physically possible \cite{palmerend}. Therefore, the distribution of $\hat{P}$ can be considered to be continuous. Let us assume, for simplicity, that $\hat{p}$ is constant for all the runs. The distribution of $\hat{P}$ depends on $\hat{p}$ as $|\hat{P} - \hat{p}|< \Delta$. The distribution may also be correlated with the experimenter's choice of the preparation setting $\hat{a}$. Therefore, we define a (normalised) continuous distribution $\rho(\hat{P}|\hat{a}, \hat{p})$. The distribution $\rho(\hat{P}|\hat{a}, \hat{p}) >0$ only if $|\hat{P}| = 1$. Let $\hat{p}'$ label the average initial exact orientation of the preparation apparatus over an ensemble of runs. That is,
\begin{align}
\hat{p}' = \int \textit{ }\rho(\hat{P}|\hat{a}, \hat{p}) \hat{P} d\Omega_P 
\end{align}
This implies that
\begin{align}
|\hat{p}' - \hat{p}| &= |\int \textit{ }\rho(\hat{P}|\hat{a}, \hat{p}) \hat{P} d\Omega_P - \hat{p}| \\
&= |\int \textit{ }\rho(\hat{P}|\hat{a}, \hat{p}) (\hat{P} - \hat{p}) d\Omega_P |\\
&\leq \int \textit{ }\rho(\hat{P}|\hat{a}, \hat{p}) |\hat{P} - \hat{p}| d\Omega_P 
\end{align}
As $|\hat{P} - \hat{p}| < \Delta$, 
\begin{align}
|\hat{p}' - \hat{p}| < \int \textit{ }\rho(\hat{P}|\hat{a}, \hat{p}) \Delta d\Omega_P  = \Delta \label{0}
\end{align}

Equation (\ref{0}) implies that the experimenter cannot distinguish between the initial orientation that they have chosen and the average initial exact orientation over the ensemble, due to the finite resolution of the apparatus used for setup. The average value of $\hat{A}$ will be
\begin{align}
\hat{a}' \equiv \int \textit{ }\rho(\hat{P}|\hat{a}, \hat{p})  \hat{A}(\hat{P}, \hat{p}, \hat{a}) d\Omega_P \label{behenka}
\end{align}
Using the fact that $|\hat{A}(\hat{P}, \hat{p}, \hat{a}) - \hat{a}|< \Delta$, one can show that $|\hat{a}' - \hat{a}|< \Delta$, analogous to equation (\ref{0}). This implies that, for an ensemble of runs, the average final exact orientation of the preparation apparatus cannot be experimentally distinguished from the experimentally-selected orientation $\hat{a}$.\\

\begin{figure}
\includegraphics[scale=0.35]{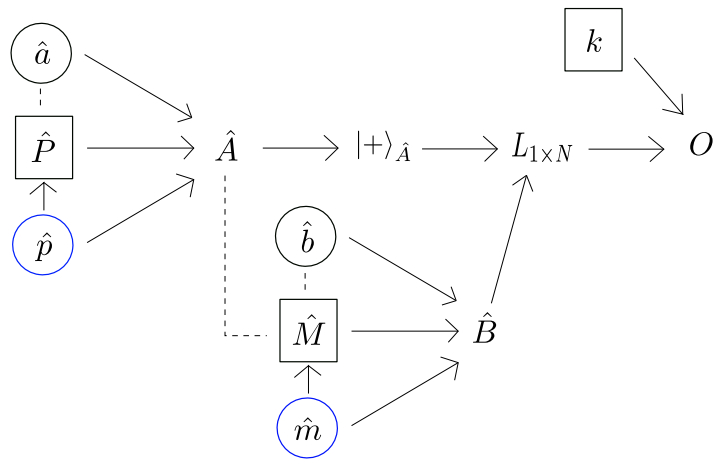}
\caption{Schematic illustration of the relationships between the experimenters' choices, hidden variables and the measurement outcome in the model. Causal influences are depicted by a directed arrow, whereas correlations are depicted via a dashed line. The experimenter's choices $\hat{p}$, $\hat{a}$ ($\hat{m}$, $\hat{b}$) at the preparation (measurement) end are circled in the figure. The variables $\hat{p}$, $\hat{m}$, circled in blue, are considered to be constant for all the runs whereas $\hat{a}$, $\hat{b}$ vary with each run in general. At the preparation (measurement) end, the hidden variable $\hat{P}$ ($\hat{M}$) is causally influenced by $\hat{p}$ ($\hat{m}$) and correlated with $\hat{a}$ ($\hat{b}$). The hidden variable $\hat{M}$ is also correlated with $\hat{A}$ due to the rationality constraint (\ref{a}). The experimenter's choices $\hat{p}$, $\hat{a}$ ($\hat{m}$, $\hat{b}$) and the hidden variable $\hat{P}$ ($\hat{M}$) jointly determine the final exact preparation (measurement) setting $\hat{A}$ ($\hat{B}$). The variable $\hat{A}$ determines the exact quantum state $|+\rangle_{\hat{A}}$ prepared, and $|+\rangle_{\hat{A}}$, $\hat{B}$ in turn jointly determine the bit string $L_{1 \times N}$. The measurement outcome $O$ is determined by $L_{1 \times N}$ and the hidden variable $k$. The variables $\hat{P}$, $\hat{M}$, $k$, $\hat{A}$, $\hat{B}$, $|+\rangle_{\hat{A}}$ and $L_{1 \times N}$ are hidden variables that determine the measurement outcome. The hidden variables $\hat{P}$, $\hat{M}$, $k$ do not causally depend on the experimenter's choices $\hat{a}$, $\hat{b}$ and are boxed. They constitute the hidden variable $\mu \equiv (\hat{P}, \hat{M}, k$) which remains constant during a counter-factual change of $\hat{a}$, $\hat{b}$ for a particular run.}
\end{figure}

Let us now describe the distribution of $\hat{M}$ at the measurement end. We assume, for simplicity, that $\hat{m}$ is constant for all the runs. Consider the distribution of the initial exact apparatus orientation $\hat{M}$. This distribution depends non trivially on the variable $\hat{A}$ at the preparation end due to equation (\ref{a}), which implies that, given an orientation $\hat{A}$, the distribution of $\hat{B}$ must be discrete. As $\hat{B} = \hat{B}( \hat{M}, \hat{m}, \hat{b})$, it is natural to consider $\hat{M}$ to be discrete as well\footnote{It is possible to consider $\hat{M}$ as continuous, but this complicates the mathematical analysis without leading to any physical insight.}. Suppose that the experimentally-selected final orientation of the apparatus for a particular run is $\hat{b}$. Then, one can define a discrete distribution $p(\hat{M}_i| \hat{A}, \hat{b}, \hat{m})$ such that $\sum_i p(\hat{M}_i| \hat{A}, \hat{b}, \hat{m}) = 1$, and $p(\hat{M}_i| \hat{A}, \hat{b}, \hat{m}) > 0$ only if $|\hat{M}_i - \hat{m}| < \Delta$ and $\hat{B}(\hat{M}_i, \hat{m}, \hat{b})$ satisfies the constraint (\ref{a}) for some $n$. Let
\begin{align}
\hat{m}' \equiv \sum_i p(\hat{M}_i| \hat{A}, \hat{b}, \hat{m}) \hat{M}_i 
\end{align}
One can prove, analogous to equation (\ref{0}), that $| \hat{m}' - \hat{m}| < \Delta$ by using the fact that $|\hat{M}_i - \hat{m}| < \Delta$. The average value of $\hat{B}$ will be 
\begin{align}
\hat{b}' \equiv \sum_i p(\hat{M}_i| \hat{A}, \hat{b}, \hat{m}) \hat{B}(\hat{M}_i, \hat{m}, \hat{b}) \label{baap}
\end{align}
Using the fact that $|\hat{B}_i - \hat{b}| < \Delta$, where $\hat{B}_{i} = \hat{B}( \hat{M}_i, \hat{m}, \hat{b})$, one can further show that $|\hat{b}' - \hat{b}| < \Delta$. Lastly, the Proposal defines the distribution of $k$ over the $N$ values to be uniform. That is, $p(k) = 1/N \textbf{ }\forall k$. \\

We now prove that the model, as defined above, reproduces the quantum predictions. We assume, for simplicity, that the initial experimentally-selected orientation of the preparation (measurement) apparatus is $\hat{p}$ ($\hat{m}$) for all runs. Let the final experimentally-selected orientations of the preparation and measurement apparatuses be $\hat{a}$ and $\hat{b}$ respectively for all the runs. The model predicts the following expectation value of outcomes
\begin{align}
& \sum_{k=1}^{N} \sum_{i}^{\alpha} \int \rho(\hat{P}| \hat{p}, \hat{a}) p(\hat{M}_i| \hat{A}, \hat{b}, \hat{m})p(k)  O\big (\hat{A}(\hat{P}, \hat{p}, \hat{a}), \hat{B}(\hat{M}_i, \hat{m}, \hat{b)}, k\big ) \textbf{ } d\Omega_P\\
&= \sum_{i}^{\alpha} \int \rho(\hat{P}| \hat{p}, \hat{a}) p(\hat{M}_i| \hat{A}, \hat{b}, \hat{m}) \hat{A}(\hat{P}, \hat{p}, \hat{a})\cdot \hat{B}_{i} \textbf{ } d\Omega_P
\end{align}
Using equations (\ref{behenka}) and (\ref{baap}), this can be simplified to
\begin{align}
& \big (\int  \rho(\hat{A}| \hat{p}, \hat{a}) \hat{A} \cdot ( \sum_{i}^{\alpha} p(\hat{M}_i| \hat{b}, \hat{m}, \hat{A}) \hat{B}_i )\textbf{ }d\Omega_A \big ) \\
&= \hat{a}'\cdot \hat{b}'
\end{align}
On the other hand, orthodox quantum mechanics predicts the expectation value to be $_{\hat{a}}\langle +| \hat{\sigma}_{\hat{b}}| +\rangle_{\hat{a}} = \hat{a}\cdot \hat{b}$. As $|\hat{a}' - \hat{a}| < \Delta$ and $|\hat{b}' - \hat{b}|< \Delta$, this difference in prediction cannot, in principle, be experimentally detected due to the finite resolution of the experimental apparatuses. Thus, the model may be said to experimentally reproduce the quantum predictions for a single spin-1/2 particle.

\section{Discussion of the single-particle model}\label{r3}
In this section, we use our model to analyse the different arguments in the Proposal about single spin-1/2 particles. We will show that several of these arguments, that appear reasonable in the Proposal, do not work in our model. We begin with a discussion of some properties.\\

\subsection{Properties of the model} \label{r3a}
1. Measurement dependence: A hidden-variable model is called measurement dependent if the hidden-variable distribution is correlated with the measurement settings. In the Proposal, the exact measurement settings and the experimentally-selected measurement setting are different in general. Therefore, there are two possible ways to generalise the notion of measurement dependence to our model: whether the distribution of the hidden variables is correlated with $a)$ the experimentally-selected measurement settings, or $b)$ the exact measurement settings. We now prove that the model is measurement dependent according to either definition. Consider the hidden variable $\mu = (\hat{P}, \hat{M}, k)$. The distribution $p(\hat{M}| \hat{A}, \hat{b}, \hat{m})$ is correlated with the experimental measurement setting $\hat{b}$. Therefore, the model is measurement dependent in the sense of $a)$. Further, $p(\hat{M}| \hat{A}, \hat{b}, \hat{m}, \hat{B}) \neq p(\hat{M}| \hat{A}, \hat{b}, \hat{m}, \hat{B}')$ in general, as $\hat{M} = \hat{M}(\hat{B}, \hat{m}, \hat{b})$. The model is, therefore, also measurement dependent in the sense of $b)$. \\

A more intricate question is whether the model is measurement dependent if only the physically possible exact measurement settings $\big ($those that satisfy (\ref{a})$\big )$ are considered. The Proposal argues that, for these exact settings, there is no measurement dependence. We now show that this is not true for our model in general.\\

Consider that, for a particular run, the exact final setting at the preparation end is $\hat{A}$. Let us consider two exact final settings $\hat{B}_1$ and $\hat{B}_2$ at the measurement end such that $\hat{A} \cdot \hat{B}_1$ and $\hat{A} \cdot \hat{B}_2$ both satisfy constraint (\ref{a}) for the given $\hat{A}$. We first suppose that $\hat{B}_1$ and $\hat{B}_2$ are separated by a distance $> \Delta$ so that they correspond to different experimentally-selected measurement settings $\hat{b}_1$ and $\hat{b}_2$ respectively. Further suppose that $\hat{M}(\hat{B}_1, \hat{m}, \hat{b}_1) = \hat{M}(\hat{B}_2, \hat{m}, \hat{b}_2) = \hat{M}'$. We know that $p(\hat{M}'| \hat{A}, \hat{b}_1, \hat{m}) = p(\hat{B}_1 | \hat{A}, \hat{b}_1, \hat{m})$ and $p(\hat{M}'| \hat{A}, \hat{b}_2, \hat{m}) = p(\hat{B}_2| \hat{A}, \hat{b}_2, \hat{m})$. However, $p(\hat{B}_1| \hat{A}, \hat{b}_1, \hat{m}) \neq p(\hat{B}_2| \hat{A}, \hat{b}_2, \hat{m})$ in general, which implies that $p(\hat{M}'| \hat{A}, \hat{b}_1, \hat{m}) \neq p(\hat{M}'| \hat{A}, \hat{b}_2, \hat{m})$ in general. Therefore, measurement independence is violated in the sense of $a)$ in general. Second, suppose that $\hat{B}_1$ and $\hat{B}_2$ are separated by a distance $< \Delta$ so that they correspond to the same experimentally set measurement setting $\hat{b}$. As $\hat{M}(\hat{B}_1, \hat{m}, \hat{b}) \neq \hat{M}(\hat{B}_2, \hat{m}, \hat{b})$, $p\big (\hat{M}(\hat{B}_1, \hat{m}, \hat{b})| \hat{A}, \hat{b}, \hat{m} \big ) \neq p\big (\hat{M}(\hat{B}_2, \hat{m}, \hat{b})| \hat{A}, \hat{b}, \hat{m}\big )$ in general. Therefore, measurement independence is violated in the sense of $b)$ as well in general. To summarise, the model is measurement dependent even if only the exact measurement settings that satisfy (\ref{a}) are considered.\\

2. Reality of the quantum state: In the ontological models framework \cite{harrikens, pbr, leifer}, a model is $\psi$-ontic if different quantum states have disjoint supports over the hidden-variable space. We have discussed in the previous section that the hidden variables encode the actual (exact) prepared quantum state for each run of the experiment. This implies that, given two different quantum states, their supports will be disjoint. Therefore, our model is $\psi$-ontic. \\

One may ask whether the definition of $\psi$-onticity is applicable to our model since it violates measurement independence, which is assumed in the ontological models framework. In general, extra analysis (beyond that of ontological models framework) may be required for a model that violates measurement independence. For example, a retrocausal model may be classified as $\psi$-ontic even if the prepared quantum state is epistemic given a backwards (in time) evolving quantum state in the model that is ontic, on grounds of time symmetry. However, our model posits only a forwards-evolving prepared quantum state, so that the question of $\psi$-onticity reduces to whether this prepared quantum state is real. As the exact prepared quantum state is encoded in the hidden variables, the exact prepared quantum state is real and the model is $\psi$-ontic. \\

For an intuitive understanding of the reality of the exact quantum state, let us consider the interpretation that the exact quantum state is not a hidden variable, but an epistemic variable that determines probabilities at the ensemble level. We can easily show that this interpretation is incorrect. Firstly, we know that the outcome of an individual run $O(\mu, \hat{p}, \hat{a}, \hat{m}, \hat{b}) = L_{1, k} (|+\rangle_{\hat{A}}, \hat{B})$ is a function of the exact quantum state. This shows that the exact prepared quantum state plays a role in each run of the experiment, not just at the ensemble level. Secondly, the exact quantum state varies with each run in general. But the exact quantum state must be constant over the runs if it is supposed to represent an ensemble of runs. Therefore, we conclude that the exact quantum state is a hidden variable in our model.\\

We note that the quantum state $|+\rangle_{\hat{a}}$ may be considered to be an epistemic variable as it determines the outcome probabilities for an ensemble, and it depends on the experimenter's (incomplete) information about the preparation procedure. This does not, however, render the model $\psi$-epistemic as the quantum state actually prepared ($|+\rangle_{\hat{A}}$) is a hidden variable. This is consistent with the discussion of $\psi$-onticity in other models. For example, one may introduce apparatus errors in pilot-wave theory \cite{bohm1, bohm2, solventini}, but pilot-wave theory is $\psi$-ontic as the quantum state actually prepared is a hidden variable.\\

Lastly, we note that the exact quantum state $|+\rangle_{\hat{A}}$ is not included in $\mu = (\hat{P}, \hat{M}, k)$, but $\mu$ contains only those hidden variables that remain constant during a counter-factual change of the experimenter's choices $\hat{a}$, $\hat{b}$ (see section \ref{r2}).\\

3. Role of the bit string: The Proposal has argued that the bit string represents an ensemble of runs. However, the bit string $L_{1 \times N} (|+\rangle_{\hat{A}}, \hat{B})$ varies with each run, so this interpretation cannot be correct. Furthermore, all the entries of the bit string are determined by the final exact apparatus settings at the preparation end $\hat{A}$ and at the measurement end $\hat{B}$. Therefore, the bit string is real as all of its entries are encoded in the hidden variables. \\

For a particular run, the $k^{th}$ entry of the bit string for that run specifies the measurement outcome in our model. On the other hand, the other entries of the bit string for that run play no role in determining the measurement outcome. Therefore, the bit string contains redundant information in the model.

\begin{figure}
\includegraphics[scale=0.4]{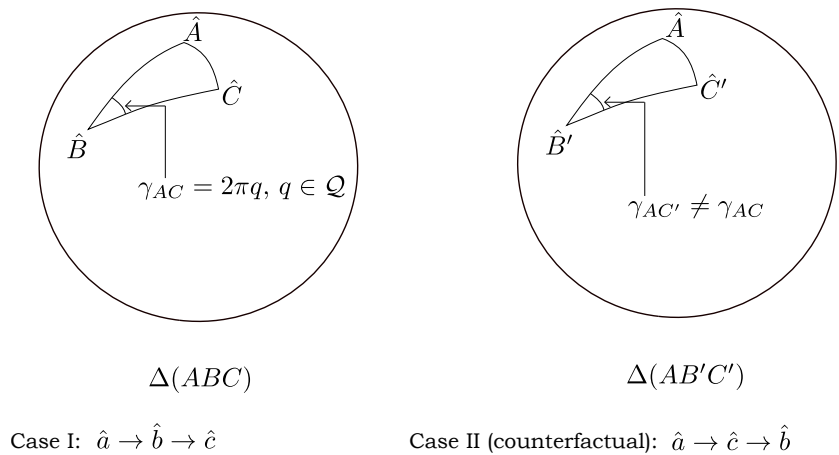}
\caption{Schematic illustration of a counterfactual change of the order of experimentally-selected settings in sequential Stern-Gerlach measurements. The left-hand side of the figure depicts a scenario where the experimenters have chosen the experimentally-selected settings to be $\hat{a} \to \hat{b} \to \hat{c}$ for a particular run of a sequential Stern-Gerlach measurement. The corresponding exact measurement settings $\hat{A}$, $\hat{B}$ and $\hat{C}$ for that run form the spherical triangle $\Delta(ABC)$ on the unit sphere. The proposal requires that $\gamma_{AC}/2\pi \in \mathcal{Q}$, where $\mathcal{Q}$ is the set of rational numbers. This implies, via Niven's theorem \cite{niven}, that $\hat{A}\cdot \hat{C} \not \in \mathcal{Q}$ (except for certain special cases). The right-hand side of the figure depicts a counter-factual scenario for that run, where the order of experimentally-selected settings is $\hat{a} \to \hat{c} \to \hat{b}$, and the times at which the measurements occur are the same as before. For this counter-factual scenario, the exact measurement settings turn out to be $\hat{A}$, $\hat{B}'$ ($ \neq \hat{B}$ in general) and $\hat{C}'$ ($\neq \hat{C}$ in general) in the model, and the relevant spherical triangle is $\Delta(AB'C')$. For the $\Delta(AB'C')$,  $\gamma_{AC'}/2\pi \not \in \mathcal{Q}$ in general and therefore it is possible that $\hat{A}\cdot \hat{C}' \in \mathcal{Q}$. Whether such a counter-factual scenario could have occurred for that run depends on whether $\gamma_{AC'}/2\pi$ is rational.}
\end{figure}

\subsection{Sequential Stern-Gerlach measurements}\label{ola}
Consider an individual run of an experiment where three sequential Stern-Gerlach measurements are performed on an ensemble of spin-1/2 particles. Let the initial experimentally-selected orientation of the first Stern Gerlach be $\hat{m}_1$ and the experimenter, for this run, choose the final orientation to be $\hat{a}$. Similarly, let the initial experimentally-selected orientation of the second (third) Stern Gerlach be $\hat{m}_2$ ($\hat{m}_3$) and the experimenter, for this run, choose the final orientation to be $\hat{b}$ ($\hat{c}$). \\ 

We know that the exact initial apparatus orientations are hidden variables that vary from one run to the next in our model. For this run, let the first, second and third apparatuses have the exact initial orientations $\hat{M}_1$, $\hat{M}_2$ and $\hat{M}_3$ respectively. The final exact orientations will then be
\begin{align}
&\hat{A} = \hat{A}(\hat{M}_1, \hat{m}_1, \hat{a}) \label{c}\\ 
&\hat{B} = \hat{B}(\hat{M}_2, \hat{m}_2, \hat{b})\label{z}\\
&\hat{C} = \hat{C}(\hat{M}_3, \hat{m}_3, \hat{c})\label{w}
\end{align}
We also know, from the previous section, that the final exact orientations must satisfy the constraints
\begin{align}
\hat{A}\cdot \hat{B} = 1 - \frac{2n_{AB}-1}{N/2} \label{co1}\\
\hat{B}\cdot \hat{C} = 1 - \frac{2n_{BC}-1}{N/2}\label{co2}
\end{align}
for some $n_{AB}, n_{BC} \in \{1, 2,....N/2 \}$ to be physically possible. The question the Proposal raises is whether, for this very run, one could have performed the measurements in the order $\hat{a} \rightarrow \hat{c} \rightarrow \hat{b}$. The Proposal argues that this is impossible. It first assumes that the sequence of exact settings is changed from $\hat{A} \rightarrow \hat{B} \rightarrow \hat{C}$ to $\hat{A} \rightarrow \hat{C} \rightarrow \hat{B}$, then, by using the constraints (\ref{co1}) and (\ref{co2}), rules out the possibility of such a change. We now show that this argument fails in our model of the Proposal.\\

Let us consider the different physical procedures by which one may change the order of final experimentally-selected orientations. First, one can switch the orientations of the apparatuses. That is, the experimentally-selected orientation of the second (third) apparatus can be changed to $\hat{c}$ ($\hat{b}$) instead of $\hat{b}$ ($\hat{c}$). Alternatively, one can change the order of apparatuses itself, while keeping their orientations fixed. That is, one can use the third apparatus (with the experimentally-selected orientation $\hat{c}$) before the second apparatus (with the experimentally-selected orientation $\hat{b}$). Let us consider both the possibilities individually:\\

a) Changing the orientations of apparatuses: Here we make use of the fact that the exact initial apparatus orientations, being causally unaffected by the experimenter's choice of $\hat{a}$, $\hat{b}$ and $\hat{c}$, will remain the same. Therefore, for this counter-factual scenario the new exact settings will be 
\begin{align}
\hat{A} = & \textbf{ }\hat{A}(\hat{M}_1, \hat{m}_1, \hat{a})\\
\hat{C}' = & \textbf{ }\hat{C}'(\hat{M}_2, \hat{m}_2, \hat{c})\\
\hat{B}' = & \textbf{ }\hat{B}'(\hat{M}_3, \hat{m}_3, \hat{b})
\end{align}
It is clear that the new exact orientations $\hat{B}'$ and $\hat{C}'$ will be different in general compared to the original exact orientations $\hat{B}$ and $\hat{C}$. \\

b) Changing the order of apparatuses: Here we make use of the fact that the exact initial apparatus orientations are continuously varying with time (see section \ref{r1}). If the ordering is changed, then the apparatuses will get used at different times than previously. That is, if the second (third) measurement occurred at time $t_2$ ($t_3$), then in the counterfactual scenario the second (third) apparatus will be used at time $t_3$ ($t_2$). Therefore, the exact initial apparatus orientations will also change. Let the exact initial orientation of the second apparatus change from $\hat{M}_2 (t_2) \rightarrow \hat{M}_2 (t_3) \equiv \hat{M}'_2$ and that of the third apparatus change from $\hat{M}_3 (t_3) \rightarrow \hat{M}_3 (t_2) \equiv \hat{M}'_3$ due to this change in order. The new exact settings will then be
\begin{align}
\hat{A} = & \textbf{ }\hat{A}(\hat{M}_1, \hat{m}_1, \hat{a})\\
\hat{C}'' = & \textbf{ }\hat{C}''(\hat{M}'_3, \hat{m}_3, \hat{c})\\
\hat{B}'' = & \textbf{ }\hat{B}''(\hat{M}'_2, \hat{m}_2, \hat{b}) 
\end{align}
Again, it is clear that the new exact final orientations $\hat{B}''$ and $\hat{C}''$ will be different in general compared to the original exact final orientations $\hat{B}$ and $\hat{C}$.\\

Therefore, we see that in our model of the Proposal, interchanging the order of experimentally-selected orientations $\hat{b}$ and $\hat{c}$ for a particular run results in different exact orientations than previously. Whether such a change is possible depends, then, on whether the new exact orientations satisfy the constraints (\ref{co1}) and (\ref{co2}). It is incorrect to rule out the possibility of such a change by assuming the exact orientations to be the same as before. We illustrate this point in Fig. 2.\\

In the next subsection, we show how a similar analysis leads to a failure of the Proposal's argument about non-commutativity of quantum observables.\\

\subsection{Non-commutativity of quantum observables}
Consider a particular run of a Stern-Gerlach measurement on an ensemble of spin-1/2 particles, for which the experimentally-selected (exact) orientation of the Stern Gerlach is $\hat{a}$ ($\hat{A}$). The Proposal shows that, for any three mutually orthogonal directions $\{\hat{X}_1, \hat{X}_2, \hat{X}_3\}$, if $\hat{A}\cdot \hat{X}_1$ satisfies the constraint (\ref{a}), then $\hat{A}\cdot \hat{X}_2$ and $\hat{A}\cdot \hat{X}_3$ do not satisfy the constraint (and so on for all other permutations). Therefore, for a particular run (with a fixed $\hat{A}$), only one of the three measurements $\{\hat{X}_1,\hat{X}_2, \hat{X}_3\}$ is well defined. The Proposal argues that the non-commutativity of quantum observables (for example $\hat{\sigma}_x$, $\hat{\sigma}_y$ and $\hat{\sigma}_z$) is thereby naturally obtained as a consequence of the constraint (\ref{a}). This is, however, not a satisfactory argument. Consider a fourth direction $\hat{X}_4 = \cos \theta \hat{X}_1 + \sin \theta (\cos \phi \hat{X}_2 + \sin \phi \hat{X}_3)$ that is non-orthogonal to $\hat{X}_1,\hat{X}_2$ and $\hat{X}_3$. According to orthodox quantum mechanics, the observable $\hat{\sigma}\cdot\hat{X}_4$ is non-commutating with all the three observables $\hat{\sigma}\cdot \hat{X}_i$ in general, where $i \in \{1, 2, 3\}$. However, both $\hat{A}\cdot \hat{X}_4$ and one of $\hat{A}\cdot \hat{X}_i$ may satisfy the constraint (\ref{a}). Therefore, the rationality constraint (\ref{a}) does not explain the non-commutativity of all quantum observables. Further, in our model of the Proposal, it is an artificial assumption to consider three mutually orthogonal directions $\{\hat{X}_1, \hat{X}_2, \hat{X}_3\}$ as the possible exact measurement settings for any run. This is because the experimenter does not have sufficient control over the exact orientations of the Stern Gerlach to ensure mutual orthogonality. Say the experimenter decides to choose from three mutually orthogonal measurement settings $\{\hat{x}_1, \hat{x}_2, \hat{x}_3\}$. Let the exact initial orientation of the Stern Gerlach for a particular run be $\hat{M}$. The possible exact final orientations for that run will be 
\begin{align}
\hat{X}_1 =  \hat{X}_1(\hat{x}_1, \hat{M}, \hat{m}) \nonumber\\
\hat{X}_2 =  \hat{X}_2(\hat{x}_2, \hat{M}, \hat{m}) \label{gao}\\
\hat{X}_3 =  \hat{X}_3(\hat{x}_3, \hat{M}, \hat{m}) \nonumber
\end{align}
It is clear from equations (\ref{gao}) that the exact measurement settings will not be mutually orthogonal in general. To summarise, it is possible for the experimenter to choose from three mutually orthogonal measurement settings $\{\hat{x}_1, \hat{x}_2, \hat{x}_3\}$ in our model of the Proposal, as the corresponding exact measurement settings are not mutually orthogonal in general.

\section{Extension of the model to the Bell scenario}\label{r4}
The model in section \ref{r2} can be easily generalised to multiple particles as we have identified the properties of the model for a single particle. Consider the standard Bell scenario \cite{bell}, where two spin-1/2 particles prepared in the spin-singlet state are subjected to local spin measurements in a space-like separated manner. We assume, for simplicity, that the initial (final) experimentally-selected orientations are $\hat{m}_1$ ($\hat{b}$) and $\hat{m}_2$ ($\hat{c}$) at wings 1 and 2 respectively for all the runs. For a particular run, let the exact initial orientation of the apparatus at wing 1 (2) be $\hat{M}_1$ $\big ( \hat{M}_2 \big )$. We assume that the final exact orientations, analogous to the single-particle case, are functions of the initial exact orientations and the experimental choices. The exact final orientation at wing 1 (2) can then be expressed as $\hat{B} = \hat{B}( \hat{M}_1, \hat{m}_1, \hat{b})$ $\big (\hat{C}= \hat{C}( \hat{M}_2, \hat{m}_2, \hat{c}) \big )$ for that run. \\

Let us describe the ontology of the model. In the single-particle case, the exact prepared quantum state $|+\rangle_{\hat{A}}$ is a hidden variable that is determined by $\hat{P}$ and the experimentally-selected orientations $\hat{p}$ and $\hat{a}$. For the Bell scenario, we analogously define $|\psi\rangle_{singlet}$ as a hidden variable. Note that the analogy is not exact as $|+\rangle_{\hat{A}}$ varies with each run unlike $|\psi\rangle_{singlet}$, but we follow the Proposal in not making a distinction between the exact quantum state and the experimentally-selected quantum state in the Bell scenario \cite{palmerend}. It appears likely that the model can be straightforwardly generalised to include an exact quantum state that is different from the experimentally-selected quantum state in the Bell scenario, but then our model would not be an accurate representation of the Proposal. For the single-particle case, the exact initial apparatus orientations $\hat{P}$ and $\hat{M}$ are treated as hidden variables. For the Bell scenario, we correspondingly define the exact initial apparatus orientations $\hat{M}_1$ and $\hat{M}_2$ at wings 1 and 2 respectively as hidden variables. The hidden variable $k$ is still the same; that is, $k \in \{1, 2, ...N\}$ where $N$ is very large even constant in the model. We define $\mu \equiv \big ( |\psi\rangle_{singlet}, \hat{M}_1, \hat{M}_2, k \big )$. As in the single-particle case, $\mu$ contains all the hidden variables that remain constant during a counter-factual change of the experimenters' measurement settings $\hat{b}$ and $\hat{c}$.\\

\begin{figure}
\includegraphics[scale=0.4]{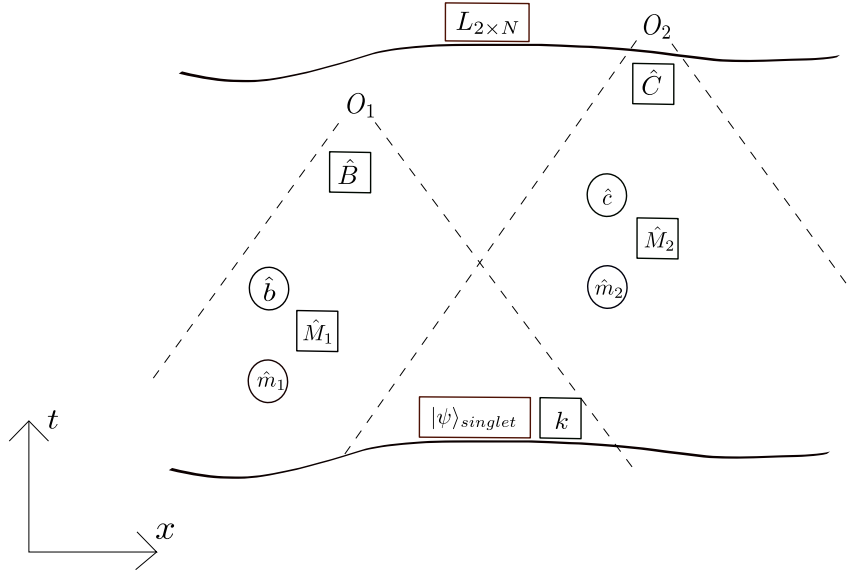}
\caption{Schematic space-time diagram of the Bell-scenario model. The prepared quantum state $|\psi(t)\rangle_{singlet}$ is defined on spacelike hypersurfaces belonging to the spacetime foliation corresponding to $t$. The final exact orientation of the measuring apparatus at wing 1 (2) $\hat{B}$ ($\hat{C}$) is determined by the initial exact orientation $\hat{M}_1$ ($\hat{M}_2$) and the local experimental choices $\hat{m}_1$ ($\hat{m}_2$) and $\hat{b}$ ($\hat{c}$). The initial exact orientation at wing 1 (2) $\hat{M}_1$ ($\hat{M}_2$) is causally influenced by the initial experimental choice $\hat{m}_1$ ($\hat{m}_2$) and correlated with the final experimental choice $\hat{b}$ ($\hat{c}$) at wing 1 (2). The variable $k$ determines the position or ``trajectory'' \cite{palmerend} in the bit string $L_{2 \times N}$, whose elements in row 1 are determined by $|\psi\rangle_{singlet}$ and  elements in row 2 are determined by $|\psi\rangle_{singlet}$ and $\hat{B}$, $\hat{C}$. The bit string $L_{2 \times N}$ and $k$ are defined on the spacelike hypersurfaces associated with $|\psi(t)\rangle_{singlet}$. The outcome at wing 1 $O_1$ is given by $L_{1, k}(|\psi\rangle_{singlet})$. The outcome at wing 2 $O_2$ is given by $L_{2, k} (|\psi\rangle_{singlet}, \hat{B}, \hat{C})$. We box (circle) the hidden variables (experimenters' choices) in the model. }
\end{figure}

The Proposal imposes the constraint that the exact measurement settings $\hat{B}$ and $\hat{C}$ must satisfy
\begin{align}
\hat{B}\cdot \hat{C} &= 1 - \frac{4n}{N} \label{b}
\end{align}
where $n \in \{1, 2,...N/2\}$. The Proposal interprets the analogous constraint $\big ($equation (\ref{a})$\big )$ in the single-particle case to mean that the exact orientation of the measuring apparatus $\hat{B}$ has a superdeterministic dependence on the exact orientation of the preparation apparatus $\hat{A}$ in the \textit{past}. This interpretation cannot be straightforwardly applied to the Bell scenario as the measurements in the two wings occur in a space-like separated manner. Therefore, one cannot identify which measurement occurred first without a preferred foliation of space-time. However, note that $\mu$ includes the non-separable quantum state $|\psi(t)\rangle_{singlet}$. Therefore, the model implicitly contains a preferred foliation of space-time corresponding to $t$ (the frame with respect to which the quantum state is defined). Thus, it is natural to suppose that the distribution of the hidden variables $\hat{M}_1$ and $\hat{M}_2$ will depend on the time-ordering according to the preferred foliation. Suppose that the measurement at wing 1 is performed before the one at wing 2. Then, the distribution of $\hat{M}_1$ can be considered arbitrary, and the distribution of $\hat{M}_2$ will be subject to the constraint (\ref{b}). \\

For simplicity, we assume that the measurement at wing 1 occurs before the one at wing 2 for all runs. In this case, the distribution of $\hat{M}_1$ will be given by a continuous distribution $\rho(\hat{M}_1|\hat{b}, \hat{m}_1)$, where $\rho(\hat{M}_1|\hat{b}, \hat{m}_1) >0$ only if $|\hat{M}_1 - \hat{m}_1| < \Delta$. Let us label the average $\hat{M}_1$ over an ensemble of runs by $\hat{m}'_1$. That is,
\begin{align}
\hat{m}'_1 \equiv \int  \textit{ } \rho(\hat{M}_1|\hat{b}, \hat{m}_1) \hat{M}_1 d\Omega_{M_1} \label{alp}
\end{align}
We can prove that $|\hat{m}'_1 - \hat{m}_1| < \Delta$ using $|\hat{M}_1 - \hat{m}_1| < \Delta$. The average value of $\hat{B}$ will be 
\begin{align}
\hat{b}' \equiv \int \textit{ }\rho(\hat{M}_1|\hat{b}, \hat{m}_1) \hat{B}( \hat{M}_1, \hat{m}_1, \hat{b}) d\Omega_{M_1}  \label{bol}
\end{align}
Analogous to the single-particle case, we assume that $|\hat{B}( \hat{M}_1, \hat{m}_1, \hat{b}) - \hat{b}| < \Delta$ $\forall \textbf{ } \hat{M}_1, \hat{m}_1, \hat{b}$, which implies that $|\hat{b}' - \hat{b}| < \Delta$.\\

Consider the distribution of $\hat{M}_2$. Firstly, we know that $\hat{C}$ must be discrete due to the constraint (\ref{b}). Let us, analogous to the single-particle case, assume that $\hat{M}_2$ is also discrete (assuming otherwise is possible, but makes the mathematical analysis more complicated without generating physical insight). We can then define the distribution $p(\hat{M}_{2i}| \hat{B}, \hat{c}, \hat{m}_2)$ such that $\sum_i p(\hat{M}_{2i}| \hat{B}, \hat{c}, \hat{m}_2) = 1$ and $p(\hat{M}_{2i}| \hat{B}, \hat{c}, \hat{m}_2) > 0$ only if $|\hat{M}_{2i} - \hat{m}_2| < \Delta$. The average $\hat{M}_{2i}$ over an ensemble will be
\begin{align}
\hat{m}'_2 \equiv \sum_i p(\hat{M}_{2i}| \hat{B}, \hat{c}, \hat{m}_2) \hat{M}_{2i} \label{vape2}
\end{align}
We can make use of $|\hat{M}_{2i} - \hat{m}_2| < \Delta$ to prove that $|\hat{m}'_2 -\hat{m}_2| < \Delta$. Let $\hat{C}_i = \hat{C}(\hat{M}_{2i}, \hat{m}_2, \hat{c})$, where $|\hat{C}_i - \hat{c}| < \Delta$ $\forall \textbf{ } \hat{M}_{2i}, \hat{m}_2, \hat{c}$. The average value of $\hat{C}_i$ will then be
\begin{align}
\hat{c}' \equiv \sum_i p(\hat{M}_{2i}| \hat{B}, \hat{c}, \hat{m}_2) \hat{C}_i  \label{omfg2}
\end{align}
Using $|\hat{C}_i - \hat{c}| < \Delta$, we can prove that $|\hat{c}' - \hat{c}| < \Delta$. Lastly, the distribution over $k$ is assumed to be uniform as in the single-particle case. Let us now discuss the mapping from $\mu$ to the outcomes given the experimenters' choices $\hat{m}_1$, $\hat{b}$ and $\hat{m}_2$, $\hat{c}$ at the two wings.\\

For a given $\mu= \big ( |\psi\rangle_{singlet}, \hat{M}_1, \hat{M}_2, k \big )$ and the experimentally-selected orientations $\hat{m}_1$, $\hat{b}$ and $\hat{m}_2$, $\hat{c}$, The Proposal constructs the following bit-string representation of $|\psi\rangle_{singlet}$:
\begin{align}
L_{2 \times N}\big (|\psi\rangle_{singlet}, \hat{B}(\hat{M}_1, \hat{b}, \hat{m}_1), \hat{C}(\hat{M}_2, \hat{c}, \hat{m}_2)\big ) =  \begin{bmatrix} \overbrace{+1.....+1 \textbf{ }+1....+1}^{N/2} & \textbf{ }\overbrace{-1.....-1\textbf{ }-1......-1}^{N/2}  \\ \\
\underbrace{+1....+1}_{n} \textbf{ }\textbf{ }\underbrace{-1....-1}_{N/2-n} & \textbf{ }\underbrace{-1.....-1}_{n}\textbf{ }\textbf{ }\underbrace{+1.....+1}_{N/2-n} \end{bmatrix}
\end{align}
where $n$ is defined from equation (\ref{b}). The Proposal constructs bit string representation for more general entangled qubits, but we will not discuss them here. There are $N$ columns and two rows in the bit string $L_{2 \times N}(|\psi\rangle_{singlet}, \hat{B}, \hat{C})$. The elements in the upper row depend only on $|\psi\rangle_{singlet}$ and $k$, whereas the elements in the lower row are a function of $\hat{B}$, $\hat{C}$ as well as $|\psi\rangle_{singlet}$, $k$ as they depend on the variable $n$, defined in equation (\ref{b}). For a particular column $k$ in the bit string, $L_{1, k}$ ($L_{2, k}$) is defined to be the outcome at wing 1 (2). Therefore, the outcome at wing 1 may be represented by the function\footnote{It might be possible to redefine the bit string so that $O_1$ depends on the local measurement setting $\hat{B}$, but doing so would create a discrepancy between the Proposal and our model of the Proposal as the bit string $L_{2 \times N}(|\psi\rangle_{singlet}, \hat{B}, \hat{C})$ is identical to that defined by the Proposal for the singlet state \cite{palmerend}.} $O_1(\mu) = L_{1, k}(|\psi\rangle_{singlet}) $, and the outcome at wing 2 may be represented by the function $O_2(\mu, \hat{b}, \hat{m}_1, \hat{c}, \hat{m}_2) = L_{2, k} (|\psi\rangle_{singlet}, \hat{B}(\hat{M}_1, \hat{b}, \hat{m}_1), \hat{C}(\hat{M}_2, \hat{c}, \hat{m}_2))$. We also note that ($L_{1, k}$, $L_{2, k}$) is one of the following four pairs of values: $(\pm 1, \pm 1), (\pm 1, \mp 1)$. Further, there are exactly $N \frac{1 - \hat{B}\cdot \hat{C}}{4}$ columns with values $(+1, +1)$ and $(-1, -1)$ each, and there are $N\frac{1 + \hat{B}\cdot\hat{C}}{4}$ columns with values $(-1, +1)$ and $(+1, -1)$ each. Lastly, we note that both $N\frac{1 - \hat{B}\cdot\hat{C}}{4}$ and $N\frac{1 + \hat{B}\cdot\hat{C}}{4}$ are positive integers because of the constraint (\ref{b}). We now prove that the model reproduces the singlet-state correlations.\\

The model predicts the expectation value of outcomes to be
\begin{align}
& \sum_{k=1}^{N} \sum_{i}^{\alpha} \int\rho(\hat{M}_1| \hat{b}, \hat{m}_1) p(\hat{M}_{2i}| \hat{c}, \hat{m}_2, \hat{B})p(k)  O_1( \mu) 
O_2(\mu, \hat{b}, \hat{m}_1, \hat{c}, \hat{m}_2) \textbf{ } d\Omega_{M_1}\\
&= \sum_{k=1}^{N} \sum_{i}^{\alpha} \int\rho(\hat{M}_1| \hat{b}, \hat{m}_1) p(\hat{M}_{2i}| \hat{c}, \hat{m}_2, \hat{B})p(k)  L_{1, k}(|\psi\rangle_{singlet}) \nonumber \\
&\textbf{ } \textbf{ } \textbf{ } \textbf{ } \textbf{ } \textbf{ } \textbf{ } \textbf{ } \textbf{ } \textbf{ } \textbf{ } \textbf{ } \textbf{ } \textbf{ } \textbf{ } \textbf{ } \textbf{ } \textbf{ } \textbf{ } \textbf{ } L_{2, k} \big (|\psi\rangle_{singlet}, \hat{B}(\hat{M}_1, \hat{b}, \hat{m}_1), \hat{C}(\hat{M}_{2i}, \hat{c}, \hat{m}_2)\big )\textbf{ } d\Omega_{M_1}\\
&= - \sum_{i}^{\alpha} \int  \rho(\hat{M}_1| \hat{b}, \hat{m}_1) p(\hat{M}_{2i}| \hat{c}, \hat{m}_2, \hat{B}) \hat{B}(\hat{M}_1, \hat{b}, \hat{m}_1)\cdot \hat{C}_{i} \textbf{ } d\Omega_{M_1}
\end{align}
where $\hat{C}_{i} = \hat{C}(\hat{M}_{2i}, \hat{m}_2, \hat{c})$. Using the constraints (\ref{bol}) and (\ref{omfg2}), this can be simplified to
\begin{align}
& -\bigg (\int \rho(\hat{M}_1| \hat{b}, \hat{m}_1) \hat{B}(\hat{M}_1, \hat{b}, \hat{m}_1) \textbf{ } d\Omega_{M_1} \bigg ) \cdot \bigg ( \sum_{i}^{\alpha} p(\hat{M}_{2i}|\hat{B}, \hat{m}_2, \hat{c}) \hat{C}_i \bigg )\\
&= -\hat{b}'\cdot \hat{c}'
\end{align}
This cannot be experimentally distinguished from the quantum prediction $\langle\hat{\sigma}\cdot {\hat{b}} \otimes \hat{\sigma}\cdot{\hat{c}}\rangle = - \hat{b} \cdot \hat{c}$, as $|\hat{b}' -\hat{b}| < \Delta$ and $|\hat{c}' - \hat{c}| < \Delta$. Thus, the model may be said to reproduce the Bell correlations experimentally.

\section{Discussion of the Bell-scenario model}\label{r5}
In this section, we use our model for the Bell scenario to analyse several arguments in the Proposal. We begin with a discussion of some properties.

\subsection{Properties of the model}\label{r5a}
1. Measurement dependence: The distribution $\rho(\hat{M}_1|\hat{b}, \hat{m}_1)$ of the exact measurement setting at wing 1 is, in general, correlated with the local experimentally-selected measurement setting $\hat{b}$. The distribution $p(\hat{M}_{2}| \hat{B}, \hat{c}, \hat{m}_2)$ of the exact measurement setting at wing 2 is correlated with the local experimentally-selected measurement setting $\hat{c}$, and with the distant exact measurement setting $\hat{B}$ due to the constraint (\ref{b}). Therefore, the model is measurement dependent regardless of whether we define measurement dependence as correlation of hidden variables with the experimentally-selected settings or the exact settings (see section \ref{r3a}). \\

However, there remains the question whether the model is measurement dependent if only the physically possible measurement settings $\big ($those that satisfy (\ref{b})$\big )$ are considered. The Proposal argues that, if only these exact settings are considered, then there is no measurement dependence. Consider two different exact settings $\hat{B}_1$ and $\hat{B}_2$ that satisfy (\ref{b}) for a given $\hat{C}$. Then, in general
\begin{align}
p(\hat{C}| \hat{B}_1, \hat{c}, \hat{m}_2) \neq p(\hat{C}| \hat{B}_2, \hat{c}, \hat{m}_2)
\end{align}
which implies that
\begin{align}
p(\hat{M}_{2}| \hat{B}_1, \hat{c}, \hat{m}_2) \neq p(\hat{M}_2| \hat{B}_2, \hat{c}, \hat{m}_2)
\end{align}
in general. Therefore, our model of the Proposal is measurement dependent even if only the physically possible exact measurement settings are considered.\\

2. Nonlocality: The outcome at the second wing $L_{2, k} (|\psi\rangle_{singlet}, \hat{B}, \hat{C})$ is a function of the exact measurement setting $\hat{B}$ at the first wing. This implies that the probability of $O_2$ is non-trivially altered by the specification of $\hat{B}$, given complete information about the variables in the past light cone of $O_2$. As $\hat{B}$ is localised in a space-like separated region with respect to $O_2$ (see Fig. 3), the model violates local causality \cite{cuisine}. \\

Similar to the issue of measurement dependence, the Proposal argues that local causality is satisfied if only the physically possible measurement settings are considered. To see that this is not true for our model of the Proposal, consider two exact settings $\hat{B}_1$ and $\hat{B}_2$ at wing 1 that satisfy (\ref{b}) for a particular exact setting $\hat{C}$ at wing 2. There will then be two different bit-string representations of the singlet state corresponding to $\hat{B}_1$ and $\hat{B}_2$. In general, $O_2\big (L_{2 \times N}(|\psi\rangle_{singlet},\hat{B}_1, \hat{C}), k\big ) \neq O_2\big (L_{2 \times N}(|\psi\rangle_{singlet},\hat{B}_2, \hat{C}), k\big )$ as the elements will be different for the same position $k$ in the two bit strings. Thus, our model is nonlocal even if only the physically possible measurement settings are considered. 
\subsection{Counter-factual experimental settings}
The Proposal argues that it is not possible, for a particular run, to change the experimental setting at one wing without a corresponding change in the experimental setting at the other wing. The argument is as follows. Consider a particular run where the experimentally-selected measurement settings are $\hat{b}$ and $\hat{c}$ at wings 1 and 2 respectively. The corresponding exact settings $\hat{B}$ and $\hat{C}$ must satisfy the constraint (\ref{b}). Suppose the experimenter at wing 1 performs a second measurement on their particle with the experimental setting $\hat{b}'$ (corresponding to the exact setting $\hat{B}'$ for that run) after the first measurement with the experimental setting $\hat{b}$. The particle at wing 1 can then be considered to undergo a sequential Stern-Gerlach measurement, which was discussed in section \ref{ola}. Therefore, $\hat{B}$ and $\hat{B}'$ must satisfy the constraint (\ref{a}). It is then argued from the geometry of the spherical triangle $\Delta (BB'C)$ that this implies $\hat{B}'$ and $\hat{C}$ cannot satisfy the constraint (\ref{b}). From this, it is concluded that a counter-factual experimental setting $\hat{b}'$ (instead of $\hat{b}$) could not have been chosen at wing 1 as the first measurement during that run, while keeping $\hat{C}$ constant at wing 2. We show below that, in our model of the Proposal, this argument fails.\\

Consider a particular run of the experiment, with two sequential measurements occurring at wing 1. Let the initial experimentally-selected (exact) orientations of the first and the second apparatus at wing 1 be $\hat{m}_1$ ($\hat{M}_1$) and $\hat{m}'_1$ ($\hat{M}'_1$) respectively. Let the final experimentally-selected orientations of the first and the second apparatus at wing 1 be $\hat{b}$ and $\hat{b}'$ respectively. Let the exact final setting at wing 2 for that run be $\hat{C}$. We know that $\hat{B} = \hat{B}( \hat{M}_1, \hat{b}, \hat{m}_1)$ and $\hat{C}$ must satisfy the constraint (\ref{b}), and $\hat{B}$ and $\hat{B}'= \hat{B}'( \hat{M}'_1, \hat{b}', \hat{m}'_1)$ must satisfy the constraint (\ref{a}). What happens in our model if the order of final experimentally-selected orientations at wing 1 is changed from $\hat{b}\rightarrow \hat{b}'$ to $\hat{b}' \to \hat{b}$?\\

The final exact orientation of the first apparatus will then change\footnote{We assume here that the orientations of apparatuses are interchanged. It is also possible to assume, instead, that the ordering of apparatuses is interchanged while keeping their experimentally-selected orientations fixed. See section \ref{ola} for more details.} from $\hat{B}= \hat{B}( \hat{M}_1, \hat{b}, \hat{m}_1) \to \hat{B}'_0 = \hat{B}'_0( \hat{M}_1, \hat{b}', \hat{m}_1)$. Similarly, the final exact orientation of the second apparatus will change from $\hat{B}'=\hat{B}'( \hat{M}'_1, \hat{b}', \hat{m}'_1) \to \hat{B}_0 = \hat{B}_0( \hat{M}'_1, \hat{b}, \hat{m}'_1)$. Whether such a change is physically possible depends on whether $\hat{B}'_0$ and $\hat{C}$ satisfy (\ref{b}), and whether $\hat{B}'_0$ and $\hat{B}_0$ satisfy (\ref{a}). The relevant spherical triangle to consider is $\Delta(B_0'B_0C)$ -- not $\Delta (BB'C)$, as assumed in the Proposal. Therefore, the Proposal's argument that no such changes are possible, based on $\Delta (BB'C)$, is incorrect in our model of the Proposal.

\subsection{Superdeterministic conspiracy}\label{r6}
The Bell-scenario model of the Proposal illustrates a key conspiratorial feature of superdeterminism discussed in ref's \cite{1st, 2nd}. In the aforementioned references, the conspiratorial character of superdeterministic models is quantified in two separate ways. The first defines superdeterministic conspiracy in terms of a fine-tuning problem unique to superdeterministic models. The second defines it in terms of arbitrarily large correlations set up by the initial conditions. The finetuning argument cannot be directly applied to our model as initial conditions that lead to exact apparatus orientations that violate the rationality constraints (\ref{a}) and (\ref{b}) are considered to be unphysical by the Proposal. However, the second argument is readily applicable, as we show below.\\

Consider a Bell scenario in our model where there is only one Stern-Gerlach apparatus at wing 1 but $N$ apparatuses at wing 2. Let all the $N$ apparatuses at wing 2 have a common final experimentally-selected orientation for all the runs. This common orientation can vary, in general, from one run to the next. The experimenter at wing 2 can choose a different apparatus at each run to perform the measurement. Let us assume, for simplicity, that wing 1 registers an outcome before wing 2 (with respect to the foliation determined by $|\psi(t)\rangle$) for all runs of the experiment. The exact final orientation of the apparatus at wing 2 will then depend on the exact final orientation of the apparatus at wing 1 due to the constraint (\ref{b}). The question is: \textit{which} apparatus? The constraint (\ref{b}) is applicable only to the apparatus through which the quantum particle actually passes through. Therefore, the constraint will apply to different apparatuses for different runs based on the experimenter's choices. It then appears as if the experimenter's choice causally determines which apparatus will be subject to the rationality constraint (\ref{b}). However, the experimenter is only restricted by the initial conditions to choose the apparatus with the correct final exact orientation $\big ($that satisfies (\ref{b})$\big)$ for that particular run. That is, there is a one-to-one correlation -- but not causation -- between the experimenter's choice of apparatus and the apparatus in fact subject to the constraint (\ref{b}). Intuitively, one can identify this as a conspiratorial feature of the model: the experimenter does not know beforehand which apparatus will be subject to the constraint for any given run, but the initial conditions ensure that the experimenter unconsciously makes the correct choice for each run. The correlation required for this `mimicking of causation' grows with $N$ \cite{2nd}. Therefore, the initial conditions in the model must arrange arbitrarily large correlations as $N$ is increased, which is a conspiratorial feature of superdeterminism. 

\section{Conclusion}\label{r7}
The hidden-variable formulation has allowed us to make a clear assessment of several arguments made in the Proposal. We have shown that the arguments about the non-commutativity of quantum observables, the order of measurements in a sequential Stern-Gerlach measurement, and the impossibility of counterfactual measurements in Bell experiments fail in our model of the Proposal. All three arguments have been undermined by a proper consideration of the exact orientations of the measuring apparatuses. The hidden-variable model forces us to appreciate the crucial, and surprising, role played by them.\\

The hidden-variable model has also made possible a clear assessment of the properties of the Proposal. The Proposal argues that it is neither measurement dependent nor nonlocal if only the physically possible exact measurement settings are considered. However, our model for the Bell scenario is both measurement dependent and nonlocal even when so restricted. We have also shown that the model is $\psi$-ontic. The $\psi$-ontic property, in fact, turned out to be crucial in clearly defining the distribution of exact apparatus orientations in the Bell scenario. The properties of the model -- $\psi$-onticity, preferred foliation of space-time, nonlocality -- show that the relativistic motivations for considering superdeterminism are not met. Lastly, the Proposal argues that the bit string is an epistemic variable that describes ensemble probabilities. However we found that, in our model, the bit string is a hidden variable that varies with each run of the experiment. Furthermore, it contains redundant information as only its $k^{th}$ element plays a role in determining the measurement outcome. Assuming that our model is a fair representation of the Proposal, this implies a redundancy in the ideas constituting the latter. \\

We have used recent results from ref's. \cite{1st, 2nd} to quantitatively discuss the issue of superdeterministic conspiracy in the Proposal. The Proposal has argued that it involves no finetuning because the points in state-space that correspond to physically possible measurement settings are p-adic far from points that correspond to unphysical measurement settings. However, our discussion of conspiracy considers only the points that correspond to the physically possible settings, and shows that (for these points) the initial conditions must `mimick causal relationships' by arranging arbitrarily large correlations in the model. The presence of such arbitrarily large correlations, set up by the initial conditions, in a model has been argued to be a conspiratorial feature of superdeterminism \cite{1st, 2nd}.\\ 

One possibly way to circumvent our conclusions might be to claim that, despite our best efforts, the model does not accurately capture the essential ideas of the Proposal. This route is taken by Palmer and co-workers in ref. \cite{totalpagol}. Evaluating this claim opens up a question of principle: how do we know whether a model, based on a proposal, is an inaccurate representation of the proposal? The difference in clarity between a mathematical model and a proposal for a model can make any comparison a difficult task. In ref. \cite{senreply}, we suggest that this question be answered by considering the objections raised against the accuracy of the model. If the objections are incorrect, then one cannot conclude that the model is an inaccurate representation of the proposal. This is straightforward to check as as a hidden-variable model, unlike a proposal for such a model, can be transparently analysed. We show in ref. \cite{senreply} that the objections raised against the accuracy of the model are incorrect. Nevertheless, if the viewpoint that the model does not accurately represent the Proposal is taken, then the present work may be useful to further clarify the ideas contained in the Proposal, and identify the points of departure for a different hidden-variable formulation.\\

We also give a model-independent criticism of the Proposal: the rationality constraints (\ref{a}) and (\ref{b}) are artificial in the context of a physical theory. These constraints are supposed to apply to the exact measurement settings of an experiment. However no \textit{single} exact measurement setting can actually be defined for a real experiment, which always occur over a finite amount of time. We provisionally circumvented this problem for the purpose of model-building (see section \ref{r1}) by considering ideal von-Neumann measurements where the measurement interaction occurs over an infinitesimally small time interval $\delta t$. But then, this naturally precludes all real experiments.\\

To conclude, from the perspective of our model, the Proposal fails to provide a credible basis to build a superdeterministic hidden-variable account of quantum mechanics. Therefore, our work provides further support, along with the recent quantitative discussions of superdeterministic conspiracy \citep{1st, 2nd}, to the view that superdeterminism is unlikely to be the solution to the puzzle posed by the Bell correlations.

\acknowledgements
I am thankful to Antony Valentini and Matt Leifer for helpful discussions. I am also thankful to Howard Wiseman and Eric Cavalcanti for helpful comments on a previous version of the manuscript. The author was supported by a fellowship from the Grand Challenges Initiative at Chapman University.
\bibliographystyle{bhak}
\bibliography{bib}

\begin{thebibliography}{10}

\bibitem{bell}
J.~S. Bell.
\newblock {\em {Speakable and unspeakable in quantum mechanics: Collected
  papers on quantum philosophy}}.
\newblock {Cambridge Univ. Press}, {2004}.

\bibitem{louis27}
L.~de~Broglie.
\newblock {La nouvelle dynamique des quanta, in ``Cinquieme Conseil de Physique
  Solvay''(Bruxelles 1927), ed}.
\newblock {\em J. Bordet,(Gauthier-Villars, Paris)}, pages 374--406, 1928.

\bibitem{solventini}
G.~Bacciagaluppi and A.~Valentini.
\newblock {\em {Quantum theory at the crossroads: reconsidering the 1927 Solvay
  conference}}.
\newblock {Cambridge Univ. Press}, 2009.

\bibitem{bohm1}
D.~Bohm.
\newblock {A suggested interpretation of the quantum theory in terms of
  ``hidden" variables. I}.
\newblock {\em {Phys. Rev.}}, 85(2), 1952.

\bibitem{bohm2}
D.~Bohm.
\newblock {A suggested interpretation of the quantum theory in terms of
  ``hidden" variables. II}.
\newblock {\em {Phys. Rev.}}, {85}({2}), {1952}.

\bibitem{everest}
H.~Everett.
\newblock {`Relative State' Formulation of Quantum Mechanics}.
\newblock {\em Rev. Mod. Phys}, 29(3), 1957.

\bibitem{sepmolds}
L.~Vaidman.
\newblock {Many-Worlds Interpretation of Quantum Mechanics}.
\newblock In E.~N. Zalta, editor, {\em The {Stanford} Encyclopedia of
  Philosophy}. Metaphysics Research Lab, Stanford University, {F}all 2021
  edition, 2021.

\bibitem{wein09}
S.~Weinstein.
\newblock {Nonlocality without nonlocality}.
\newblock {\em Found. Phys}, 39(8):921--936, 2009.

\bibitem{sabinethink}
S.~Hossenfelder and T.~Palmer.
\newblock {Rethinking superdeterminism}.
\newblock {\em Frontiers in Physics}, 8:139, 2020.

\bibitem{whartreview}
K.~Wharton and N.~Argaman.
\newblock {Colloquium: Bell’s theorem and locally mediated reformulations of
  quantum mechanics}.
\newblock {\em Rev. Mod. Phys}, 92(2):021002, 2020.

\bibitem{hall10}
M.~J. Hall.
\newblock {Local deterministic model of singlet state correlations based on
  relaxing measurement independence}.
\newblock {\em {Phys. Rev. Lett.}}, {105}({25}), {2010}.

\bibitem{howmuch}
J.~Barrett and N.~Gisin.
\newblock {How much measurement independence is needed to demonstrate
  nonlocality?}
\newblock {\em Phys. Rev. Lett.}, 106(10):100406, 2011.

\bibitem{hall16}
M.~J. Hall.
\newblock {The significance of measurement independence for Bell inequalities
  and locality}.
\newblock In {\em {At the Frontier of Spacetime}}. {Springer}, {2016}.

\bibitem{cosmicbellII}
D.~Rauch, J.~Handsteiner, A.~Hochrainer, J.~Gallicchio, A.~S. Friedman,
  C.~Leung, B.~Liu, L.~Bulla, S.~Ecker, F.~Steinlechner, et~al.
\newblock {Cosmic bell test using random measurement settings from
  high-redshift quasars}.
\newblock {\em Phys. Rev. Lett}, 121(8), 2018.

\bibitem{hall19}
A.~S. Friedman, A.~H. Guth, M.~J. Hall, D.~I. Kaiser, and J.~Gallicchio.
\newblock {Relaxed Bell inequalities with arbitrary measurement dependence for
  each observer}.
\newblock {\em Phys. Rev. A}, 99(1):012121, 2019.

\bibitem{seneffect}
I.~Sen.
\newblock {The effect of time-dilation on Bell experiments in the retrocausal
  Brans model}.
\newblock {\em Proc. R. Soc. A}, 476(2234):20190546, 2020.

\bibitem{costacoffee}
O.~C. de~Beauregard.
\newblock {Time symmetry and the Einstein paradox}.
\newblock {\em Il Nuovo Cimento B (1971-1996)}, 42, 1977.

\bibitem{cramer}
J.~G. Cramer.
\newblock {The transactional interpretation of quantum mechanics}.
\newblock {\em {Rev. Mod. Phys}}, {58}({3}), {1986}.

\bibitem{pricebook}
H.~Price.
\newblock {\em {Time's arrow \& Archimedes' point: new directions for the
  physics of time}}.
\newblock {Oxford Univ. Press, USA}, {1997}.

\bibitem{sutherland}
R.~I. Sutherland.
\newblock {Causally symmetric Bohm model}.
\newblock {\em {Stud. Hist. Phil. Sci. B}}, {39}({4}), {2008}.

\bibitem{whartonmain}
K.~Wharton.
\newblock {Quantum states as ordinary information}.
\newblock {\em {Information}}, 5(1), 2014.

\bibitem{lazar15}
D.~Lazarovici.
\newblock {A relativistic retrocausal model violating Bell's inequality}.
\newblock {\em Proc. R. Soc. A}, 471(2173):20140454, 2015.

\bibitem{fpaper}
I.~Sen.
\newblock {A local $\psi$-epistemic retrocausal hidden-variable model of Bell
  correlations with wavefunctions in physical space}.
\newblock {\em {Found. Phys}}, 49(2), 2019.

\bibitem{dialect}
J.~S. Bell, A.~Shimony, M.~A. Horne, and J.~F. Clauser.
\newblock {An exchange on local beables}.
\newblock {\em Dialectica}, pages 85--110, 1985.

\bibitem{brans}
C.~H. Brans.
\newblock {Bell's theorem does not eliminate fully causal hidden variables}.
\newblock {\em {Int. J. Theor. Phys.}}, {27}({2}), {1988}.

\bibitem{hooft16}
G.~Hooft.
\newblock {\em The Cellular Automaton Interpretation of Quantum Mechanics}.
\newblock Springer, 2016.

\bibitem{1st}
I.~Sen and A.~Valentini.
\newblock {Superdeterministic hidden-variables models I: nonequilibrium and
  signalling}.
\newblock {\em Proc. R. Soc. A}, 476(2243):20200212, 2020.

\bibitem{2nd}
I.~Sen and A.~Valentini.
\newblock {Superdeterministic hidden-variables models II: conspiracy}.
\newblock {\em Proc. R. Soc. A}, 476(2243):20200214, 2020.

\bibitem{palmer09}
T.~Palmer.
\newblock {The Invariant Set Postulate: a new geometric framework for the
  foundations of quantum theory and the role played by gravity}.
\newblock {\em Proc. R. Soc. A}, 465(2110):3165--3185, 2009.

\bibitem{palmer1}
T.~Palmer.
\newblock {Lorenz, Godel and Penrose: new perspectives on determinism and
  causality in fundamental physics}.
\newblock {\em Contemp. Phys}, 55, 2014.

\bibitem{palmer2}
T.~Palmer.
\newblock {Bell's conspiracy, Schrodinger's black cat and global invariant
  sets}.
\newblock {\em Phil. Trans. R. Soc. A}, 373, 2015.

\bibitem{palmerend}
T.~Palmer.
\newblock {Discretization of the Bloch sphere, fractal invariant sets and
  Bell’s theorem}.
\newblock {\em Proc. R. Soc. A}, 476(2236):20190350, 2020.

\bibitem{palmerpost}
J.~R. Hance, T.~N. Palmer, and J.~Rarity.
\newblock {Experimental Tests of Invariant Set Theory}.
\newblock {\em arXiv:2102.07795}, 2021.

\bibitem{harrikens}
N.~Harrigan and R.~W. Spekkens.
\newblock {Einstein, incompleteness, and the epistemic view of quantum states}.
\newblock {\em {Found. Phys.}}, {40}({2}), {2010}.

\bibitem{pbr}
M.~F. Pusey, J.~Barrett, and T.~Rudolph.
\newblock {On the reality of the quantum state}.
\newblock {\em {Nat. Phys.}}, {8}:{475--478}, {2012}.

\bibitem{leifer}
M.~S. Leifer.
\newblock {Is the quantum state real? An extended review of $\psi $-ontology
  theorems}.
\newblock {\em {arXiv:1409.1570}}, {2014}.

\bibitem{vonN}
J.~Von~Neumann.
\newblock {\em {Mathematical foundations of quantum mechanics}}.
\newblock Princeton Univ. Press, 1955.

\bibitem{niven}
I.~Niven.
\newblock {Irrational numbers. The Carus Mathematical Monographs, No. 11},
  1956.

\bibitem{cuisine}
J.~S. Bell.
\newblock {La nouvelle cuisine}.
\newblock In {\em {Speakable and unspeakable in quantum mechanics: Collected
  papers on quantum philosophy}}. {Cambridge Univ. Press}, {2004}.

\bibitem{totalpagol}
J.~Hance, S.~Hossenfelder, and T.~Palmer.
\newblock {Comment on ``Analysis of the superdeterministic invariant-set theory
  in a hidden-variable setting"}.
\newblock {\em arXiv:2108.08144}, 2021.

\bibitem{senreply}
I.~Sen.
\newblock {Reply to superdeterminists on the hidden-variable formulation of
  Invariant-set theory}.
\newblock {\em arXiv:2109.11109}, 2021.

\end{thebibliography}

\end{document}